\documentclass[]{aa}
\usepackage{natbib}
\usepackage{longtable,lscape}
\usepackage{amsmath}
\bibpunct{(}{)}{;}{a}{}{,}
\usepackage{graphicx}

\usepackage[colorlinks=true,citecolor=blue]{hyperref}
\usepackage{color}
\usepackage{xspace}
\usepackage{dcolumn}
\usepackage{longtable}
\usepackage{lscape}
\usepackage{wasysym}
\usepackage{rotating}
\usepackage{afterpage}
\usepackage{mathtools}

\newcommand{\MJup}{M$_{\mathrm{Jup}}$\xspace}
\newcommand{\RJup}{R$_{\mathrm{Jup}}$\xspace}
\newcommand{\RSun}{R$_{\odot}$\xspace}
\newcommand{\MSun}{M$_{\odot}$\xspace}

\newcommand{\logg}{log~\emph{g}\xspace}
\newcommand{\mic}{$\mu$m\xspace}
\newcommand{\as}{\hbox{$^{\prime\prime}$}\xspace}
\def\tablebib#1{\par\vspace*{2ex}%
 \parbox{\hsize}{\leftskip0pt\rightskip0pt
 {\noindent\small\textbf{References.}~#1\par}}}

\usepackage[varg]{txfonts}
\begin{document}

%

\title{AF\,Lep\,b: the lowest mass planet detected coupling astrometric and direct imaging data \thanks{Based on observations made with the European Southern Observatory (ESO) telescopes at the
 Paranal Observatory in Chile under the program 0110.C-0233(A).}} 

\author{D. Mesa\inst{1}, R. Gratton\inst{1}, P. Kervella\inst{2},
  M. Bonavita\inst{3,1}, S. Desidera\inst{1}, V. D'Orazi\inst{4,1}, S. Marino\inst{5}, A. Zurlo\inst{6,7,8}, E. Rigliaco\inst{1}}

  \institute{\inst{1}INAF-Osservatorio Astronomico di Padova, Vicolo dell'Osservatorio 5, Padova, Italy, 35122-I \\
    \inst{2}LESIA, Observatoire de Paris, Universit\'e PSL, CNRS, Sorbonne Universit\'e, Universit\'e Paris Cit\'e, 5 place Jules Janssen, 92195 Meudon, France \\
    \inst{3}School of Physical Sciences, The Open University, Walton Hall, Milton Keynes MK7 6AA, UK \\
    \inst{4}Dipartimento di Fisica, Universit\`{a} di Roma Tor Vergata, Via della ricerca scientifica, 1 - 00133, Roma, Italy \\
    \inst{5}School of Physics and Astronomy, University of Exeter, Stocker Road, Exeter, EX4 4QL, UK \\
   \inst{6}Instituto de Estudios Astrof\'isicos, Facultad de Ingenier\'ia y Ciencias, Universidad Diego Portales, Av. Ej\'ercito 441, Santiago, Chile \\
  \inst{7}Escuela de Ingenier\'ia Industrial, Facultad de Ingenier\'ia y Ciencias, Universidad Diego Portales, Av. Ej\'ercito 441, Santiago, Chile \\
\inst{8}Millennium Nucleus on Young Exoplanets and their Moons (YEMS) \\  }

   \date{Received  / accepted }

\abstract
     {}
     {Using the direct imaging technique we searched for low mass
       companions around the star AF\,Lep that presents a significant
       proper motion anomaly (PMa) signal obtained from the comparison of Hipparcos and Gaia eDR3 catalogs.
   }
     {We observed AF\,Lep in two epochs with VLT/SPHERE using its subsystems IFS and IRDIS
       in the near-infrared (NIR) covering wavelengths ranging from the Y to
       the K spectral bands (between 0.95 and 2.3~\mic).
       The data were then reduced using the high-contrast imaging techniques
       angular differential imaging (ADI) and spectral differential
       imaging (SDI) to be able to retrieve the signal from low mass
       companions of the star.
   }
     {A faint companion was retrieved at a separation of $\sim$0.335\as from the
       star and with a position angle of $\sim$70.5$^{\circ}$ in the first epoch and with
       a similar position in the second epoch. This corresponds
       to a projected separation of $\sim$9~au. The extracted photometry allowed
       us to estimate for the companion a mass between 2 and 5~\MJup. This mass
       is in good agreement with what is expected for the dynamic mass of the
       companion deduced using astrometric measures (5.2-5.5~\MJup). This is the first companion with a mass 
       well below the deuterium burning limit discovered coupling direct imaging 
       with PMa measures. Orbit fitting done using the orvara tool allowed to further confirm
       the companion mass and to define its main orbital parameters.
   }
{}
   

\titlerunning{HIP25486}
\authorrunning{Mesa et al.}
   \maketitle
%

\section{Introduction}
\label{intro}

In the last decade or so, mainly thanks to the new generation of dedicated planet-finders like VLT/SPHERE \citep{2019A&A...631A.155B}, the Gemini Planet Imager
\citep[GPI;][]{2014PNAS..11112661M}, and Subaru/CHARIS \citep{2015SPIE.9605E..1CG}, the number of directly imaged (DI) sub-stellar companions has seen a rapid growth. 
Mainly made up of wide companions to young nearby stars, this population spans over a wide 
range of masses from planetary companions to more massive brown dwarf companions. Some 
examples of planetary mass companions are 51\,Eri\,b
\citep{2015Sci...350...64M}, HIP\,65426\,b \citep{2017A&A...605L...9C} and PDS\,70\,b 
 and c \citep{2018A&A...617A..44K,2019NatAs...3..749H}. 
The vast majority of these discoveries comes from large, blind surveys which have contributed to a first characterisation of the young sub-stellar population, highlighting a paucity of massive planets and brown dwarf companions at separations larger than 10~au \citep{2019AJ....158...13N, 2021A&A...651A..72V}, with an expected frequency of $\sim 5-6\%$. 
Blind DI surveys, like SHINE performed using VLT/SPHERE \citep{2017sf2a.conf..331C,2021A&A...651A..70D,2021A&A...651A..71L,2021A&A...651A..72V}, therefore come at a high cost in terms of telescope time, with a relatively low return in terms of new detections. 

 Our understanding of these objects thus remains severely limited by the small number of known systems. In particular, the current sparse sample of DI companions show a large diversity of spectro-photometric characteristics and orbital configurations, challenging to grasp as populations. Larger numbers of detections are hence essential to obtain a clearer picture of their formation patterns.
As a result, more recent campaigns are using careful pre-selection methods to maximise the detection yield in the shortest amount of time. Recently some success has been obtained in directly imaging 
companions previously detected through radial velocity (RV) like in the case of $\beta$\,Pic\,c 
\citep{2020A&A...642L...2N} and of HD\,206893\,c \citep{2022arXiv220804867H}. 
Other very successful programs have been the ones targeting stars showing proper motion anomalies (PMas, often referred to as $\Delta\mu$, or astrometric accelerations). These values are obtained searching for significant differences between proper motions measured over a long time baseline (e.g., Tycho-2, or Tycho Gaia Astrometric Solution - TGAS) and short-term proper motions like those measured by Hipparcos and Gaia. 
Several catalogs listing stars with PMAs obtained comparing Hipparcos data with those from both Gaia DR2 and eDR3 data are currently available \citep[see e.g.,][]{2018ApJS..239...31B,2019A&A...623A..72K,2021ApJS..254...42B,2022A&A...657A...7K} 
and have been used as starting point for new targeted surveys, which have led to the discovery of several interesting companions, including HIP~21152~B the first bound brown dwarf companion detected in the Hyades \citep{2022MNRAS.513.5588B, franson2022} and the most recent discovery of a 13-16 M$_{Jup}$ mass companion imaged around the accelerating star HIP\,99770 \citep{2022arXiv221200034C}. Recently, the results of the COPAINS survey were published by \cite{2022MNRAS.513.5588B}, showing a sub-stellar detection rate of $\sim 20\%$, thus providing a definitive proof of the efficiency of the method. \\
We exploited the catalogue presented in \citet{2022A&A...657A...7K} and obtained by comparing 
Hipparcos and Gaia eDR3 \citep{2021A&A...649A...1G} to select targets to be observed with 
VLT/SPHERE. The first selection criteria was that the PMa signal had a signal-to-noise (S/N) higher
than 3 and lower than 10 to be sure to exclude stars with signals
due to the presence of stellar mass companions. Moreover, we selected stars at distance from the 
Sun less than 50 pc to be able to image the inner part of the planetary system. Finally we 
selected stars with young ages according to literature (less than few hundreds of Myr) to be sure that the companion is bright enough to be imaged with SPHERE. \par
In this paper we present a new Jupiter analogue discovered with SPHERE around the accelerating star AF\,Lep, part of the Beta Pictoris Moving Group (BPMG). 
In Section~\ref{s:target} we present a detailed characterisation of the host star, while 
Section~\ref{s:obs} includes a description of the observations and data reduction.
The results, including a spectral and orbital characterisation of the companion, are presented in Section~\ref{s:res} and discussed in Section~\ref{s:discussion}. Finally, Section~\ref{s:conclusion} presents the study's conclusions and further implications.

\section{Host star properties}
\label{s:target}
AF\,Lep (HIP\,25486; HD\,35850) is an F8 star \citep{2006AJ....132..161G} at a distance of 26.8~pc from the Sun \citep{2022arXiv220800211G}. It is part of the BPMG \citep{2021MNRAS.502.5390P} with an estimated age of 24$\pm$3~Myr \citep{2015MNRAS.454..593B}. It hosts a debris disk with an estimated belt radius of $54\pm6$~au, but not yet resolved \citep{2021MNRAS.502.5390P}. \citet{ 2022A&A...657A...7K} found for this star a strong PMa signal with a S/N of 8.99 making it an ideal target for our program.\\
Other main characteristics of AF\,Lep, including the PMa values from \citet{2022A&A...657A...7K}, are summarised in Table~\ref{t:target}. \\
This star was classified as a spectroscopic binary (SB2 star) with a stellar companion at a separation of 0.021~au \citep{2004A&A...418..989N,2008MNRAS.389.1722E} and a mass ratio of 0.715. 
The binarity of the host star is a crucial point
to be discussed as the dynamical mass determined for a possible companion to HIP 25486 is proportional to the total mass
assumed for the star and this is different if it is a single or a multiple object. A discussion of this aspect is then crucial to interpret
the available PMa data. \par
We first notice that no indication of binarity is present in GAIA DR3 as the star has a RUWE\footnote{The Renormalised Unit Weight Error (RUWE) is a statistical indicator that can be used to assess the quality and reliability of Gaia's astrometric data. Data compatible with a single-star model will have a RUWE around 1, while a significantly higher value indicates a non-single or otherwise problematic source.} parameter of 0.918, compatible with a single source. 
Moreover, no hints of a companion are seen in Doppler imaging data by \citet{2015A&A...574A..25J}. 
In addition, a nearly equal mass binary suggested by the SB2 classification should produce a radial velocity (RV) signal with an amplitude of several km/s, while no RV variation was found by \citet{2017AJ....153..208B}, which have a precision of $\sim$200~m/s over a baseline of $\sim$20~years. 
Also, the analysis of TESS photometry shows a periodogram with a single strong peak at 1.008~days. This value is very close to the photometric period of 0.996~days given by \citet{2017A&A...607A...3M} and \citet{2015A&A...574A..25J}, with the small difference between the values likely due to aliasing. 
 The expected period for a star belonging to the BPMG is below 10~days \citep{2017A&A...607A...3M}. We would then expect that also the secondary would rotate at this rate and
 the peak should be visible in the TESS periodogram. Anyway no other peak is observed with a 
 period below 10~days aside to that due to the star. This clearly argues against the star being a SB2. 
The BVGJHK photometry of AF\,Lep is fully compatible with a star at mid between an F8V and an F9V according to \citet{2013ApJS..208....9P} and there is no indication for a luminosity excess as would be expected for an SB2 system. The values of $v~\sin{i}= 50.32$~km/s and $P_{\it rot}=0.966$~d \citep{2015A&A...574A..25J}, are reasonably compatible (though a bit slower) with the color of the star as belonging to the BPMG. 
Interferometric observations by \citet{Evans2012} exclude the presence of companions with masses as high as 75~\MJup at separation larger than 40 mas, and 33~\MJup at separation larger than 80 mas.
Moreover, given the inclination of 50 degrees \citep{2006ASPC..358..401M}, that is fully compatible with observed $v~\sin{i}$ and $P_{\rm rot}$, the RV variations seen by \citet{2004A&A...418..989N} are not compatible with a synchronized system. Finally, if the system has an inclination of 50 degrees and a mass ratio of 0.715, an amplitude of RV of $\sim 6$~km/s, as indicated by the rms of \citet{2004A&A...418..989N}, would require a semi-major axis of $\sim$13~au, that is $\sim$0.49~\as, and a period of $\sim$36~years. This can be excluded from the lack of detection in existing high-contrast imaging and would also be incompatible with the measured PMa.\\
We therefore conclude the classification as an SB2 by \citet{2004A&A...418..989N} is likely to be an artifact due to the presence of spots that altered the line profile in their observation and that AF\,Lep is a single star of spectral type between F8 and F9. 
The value of the stellar mass adopted in the rest of the discussion will therefore be of 
around 1.20~\MSun (see Table~\ref{t:target}) confirming the estimate given in \citet{2022A&A...657A...7K}.

\renewcommand{\arraystretch}{1.2}
\begin{table}
\caption{Stellar parameters of AF~Lep with references, including the HIP-EDR3 proper motion anomaly (PMa) and the corresponding tangential velocity anomaly ($dV_t$) with its position angle ($dV_t$ PA). }
\label{t:target}
\centering
\begin{tabular}{lll}
\hline \hline
Parameter    & Value          & Ref. \\
\hline
Sp.\ type    & F8~V           & 1 \\
T$_{eff}$ (K)     & 6100        & 2 \\
$\log$ g     & 4.4            & 2 \\
B-V          & 0.55           & 2,3 \\
$v \sin{i}~(km~s^{-1}$)   & 54.7 $\pm$ 0.5  & 4 \\
Mass (\MSun)        & 1.20 $\pm$ 0.06 & 5 \\
Radius (\RSun)     & 1.25 $\pm$ 0.06 & 6 \\
Period  (days)     & 0.966 $\pm$ 0.002  & 7 \\
RV (km~s$^{-1}$)& 21.10 $\pm$ 0.37 & 8\\
\hline 
     Parallax (mas) & $37.253 \pm 0.019$ & 8 \\
     Proper motion (mas~yr$^{-1}$) & $\mu_\alpha = 16.915 \pm 0.018$ & 8\\
     & $\mu_\delta = -49.318 \pm 0.016$ & \\
     \hline
     PMa (mas~yr$^{-1}$)& $\Delta \mu_\alpha = -0.206 \pm 0.021$& 9\\
     & $\Delta \mu_\delta = -0.152 \pm 0.019$ &  \\
     $dV_t$ (m~s$^{-1}$) & 32.60	$\pm$ 3.63 & 9\\
     $dV_t$ PA (\degr) & 233.50 $\pm$	4.38 & \\
\hline
Gaia Photometry: & G: 6.209 $\pm$ 0.002 & 8 \\
                 & BP: 6.488 $\pm$ 0.004 & \\
                 & RP: 5.752 $\pm$ 0.005 & \\
UBVRI Photometry & B: 6.832 $\pm$ 0.015 &  10\\
                 & V:  6.295 $\pm$ 0010 & \\

2MASS Photometry & J: 5.268 $\pm$ 0.027 & 11\\
                 & H: 5.087 $\pm$ 0.026 & \\
                 & K: 4.926 $\pm$ 0.021 & \\
\hline \hline
\end{tabular}
\tablebib{
(1)~\citet{1986AJ.....92..910E}; (2)~\citet{1994A&A...285..272T}; (3)~\citet{1996A&AS..115...41C}; (4)~(4)~\citet{2014MNRAS.444.3517M}; (5)~\citet{1996ApJ...457..340K}; (6)~\citet{1994A&A...282..899B}; (7)~\citet{2015A&A...574A..25J}; (8)~\citet{2022arXiv220800211G}; (9): {\citet{2022A&A...657A...7K}};
(10)~\citet{tycho2}; (11)~\citet{2mass}}
\end{table}
\section{Observations and data reduction}
\label{s:obs}


\begin{table*}[!htp]
  \caption{List and main characteristics of the SPHERE observations of AF\,Lep
     used for this work. }\label{t:obs}
\centering
\begin{tabular}{ccccccccc}
\hline\hline
Date  &  Obs. mode & Coronograph & DIMM seeing & $\tau_0$ & Wind speed & Field rot. & DIT & Tot. exp.\\
\hline
2022-10-16  & IRDIFS\_EXT   & N\_ALC\_YJH\_S & 0.55\as & 5.8 ms & 7.35 m/s &$13.3^{\circ}$ &  32 s &  3584 s \\
2022-12-20  & IRDIFS\_EXT   & N\_ALC\_YJH\_S & 1.05\as & 4.0 ms & 8.82 m/s &$51.8^{\circ}$ &  32 s &  3584 s \\
\hline
\end{tabular}
\end{table*}

The main characteristics of the SPHERE/SHINE observations of AF\,Lep, performed in 
October\,$16^{th}$\,2022 and December\,$20^{th}$\,2022 are summarised in Table~\ref{t:obs}.
The weather conditions were good for the first epoch, but the target was observed far from the meridian, resulting in a very low field of view (FOV) rotation angle. 
The object was observed while passing at the meridian for the second epoch, but in slightly worse weather conditions with the seeing worsening during the second part of the observation. \par
Both observations made use of the SPHERE IRDIFS\_EXT observing mode, with IFS \citep{2008SPIE.7014E..3EC} covering a spectral range between 0.95 and 1.65~\mic (Y, J, and H spectral bands) and a FOV of $1.7\as\times1.7\as$, and IRDIS \citep{2008SPIE.7014E..3LD}  operating in the K spectral band using the K12 filter pair \citep[wavelength K1=2.110~\mic; wavelength K2=2.251~\mic;][]{2010MNRAS.407...71V} on a circular FOV of $\sim$5\as. For these observations we also exploited the SPHERE adaptive optics system SAXO \citep{2006OExpr..14.7515F}. \par
Frames with satellite spots at symmetrical position with respect to the central star
were acquired before and after the science sequence to precisely define the position of the star behind the coronagraph \citep{2013aoel.confE..63L}. 
In addition, observations of the star outside the coronagraph were obtained for photometric calibration, using a neutral density filter to avoid saturation. \par
The data were reduced through the SPHERE data center \citep{2017sf2a.conf..347D}, following the data reduction and handling pipeline \citep[DRH;][]{2008ASPC..394..581P}. 
In the case of IRDIS, the required calibrations are the creation of the master dark and of the master flat-field frames and the definition of the star center. For IFS, the calibration list also includes the definition of each spectra position on the detector, the wavelength calibration, and the application of the instrumental flat that takes into account the different responses of each lenslet of the IFS array. \par
Speckle subtraction was then applied on the reduced data using both angular
differential imaging \citep[ADI;][]{2006ApJ...641..556M} and spectral differential imaging 
\citep[SDI;][]{1999PASP..111..587R} implementing both the principal components analysis 
\citep[PCA;][]{2012ApJ...755L..28S} and the TLOCI  algorithm \citep{2014IAUS..299...48M}. The application of these algorithms to the SPHERE case is described in \citet{2014A&A...572A..85Z} and \citet{2015A&A...576A.121M} and they are currently applied using the SPHERE consortium pipeline application called SpeCal \citep{2018A&A...615A..92G}.


\section{Results}
\label{s:res}


\begin{figure*}[!htp]
\centering
\includegraphics[width=0.8\textwidth]{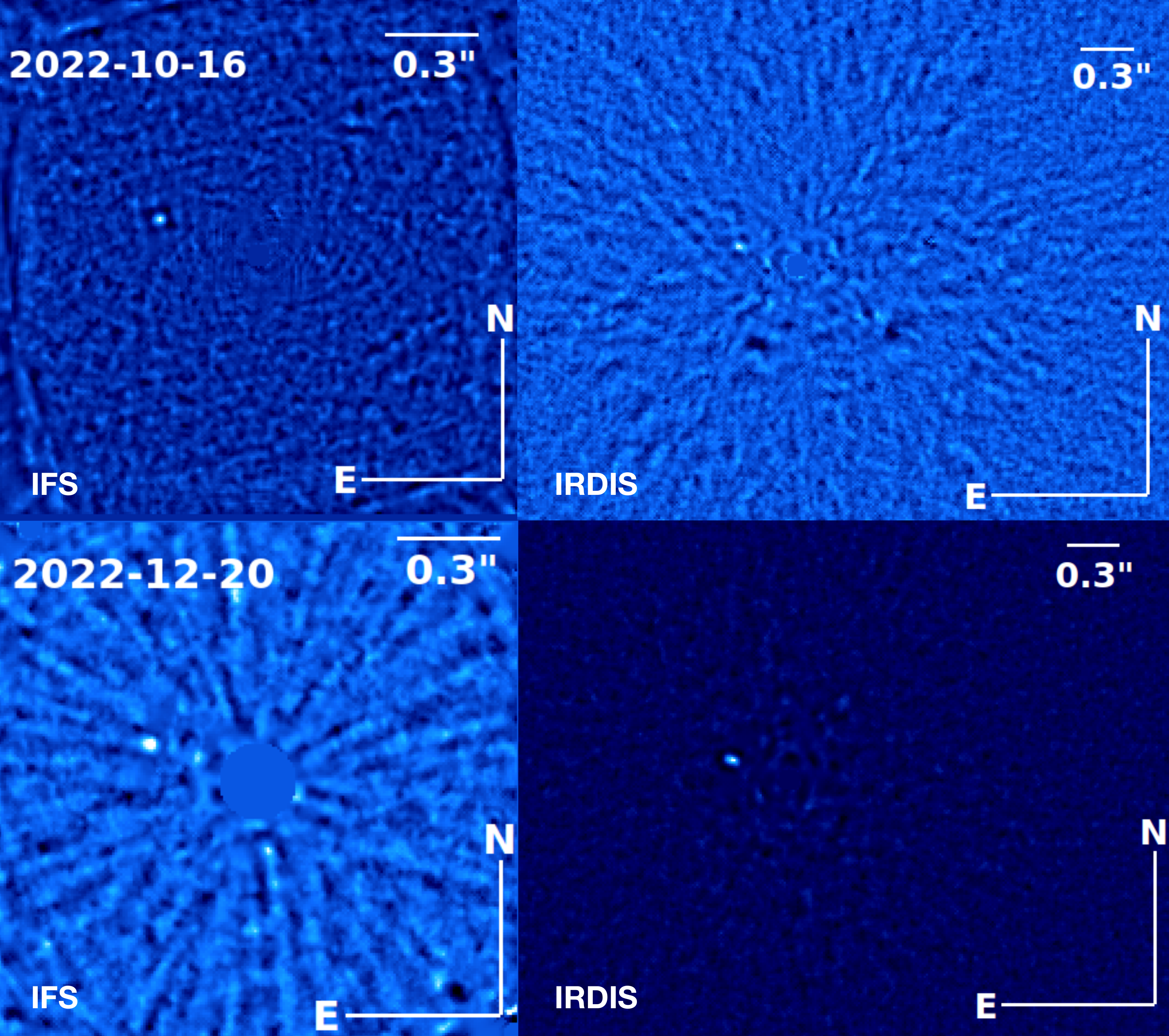}
\caption{Final S/N maps both for IFS (left panels) and IRDIS (right panels) and for both
epochs. The images for the October~$16^{th}$~2022 epoch are in the upper panels while those for the
December~$20^{th}$~2022 epoch are in the lower panels.
  For both epochs, the IFS images are obtained by applying a PCA with 150 principal components while
  in the IRDIS case, we adopted just 5 principal components. The low mass
  companion is clearly visible in all the images North-East from the star.}
\label{f:snrmap}
\end{figure*}

\begin{table*}[!htp]
  \caption{Astrometric results for AF\,Lep\,b obtained both from IFS and
  IRDIS in both observing epochs.}\label{t:astro}
\centering
\begin{tabular}{cccccccccc}
\hline\hline
 &  Date &Sep. RA  &  Err. Sep RA & Sep. Dec. & Err. Sep. Dec. & Sep. tot. & Err. Sep. tot. & PA & Err. PA \\
\hline
IFS & 2022-10-16 & 0.315\as & 0.002\as & 0.113\as & 0.001\as & 0.335\as & 0.002\as & 70.2$^{\circ}$ & 0.2$^{\circ}$ \\
IRDIS & 2022-10-16 & 0.314\as & 0.002\as & 0.110\as & 0.002\as & 0.334\as & 0.003\as & 70.7$^{\circ}$ & 0.3$^{\circ}$ \\
IFS & 2022-12-20 & 0.314\as & 0.003\as & 0.109\as & 0.004\as & 0.332\as & 0.004\as & 70.9$^{\circ}$ & 0.9$^{\circ}$ \\
IRDIS & 2022-12-20 & 0.317\as & 0.004\as & 0.110\as & 0.001\as & 0.335\as & 0.004\as & 70.8$^{\circ}$ & 0.4$^{\circ}$ \\
\hline
\end{tabular}
\end{table*}

\begin{figure}
\centering
\includegraphics[width=0.9\columnwidth]{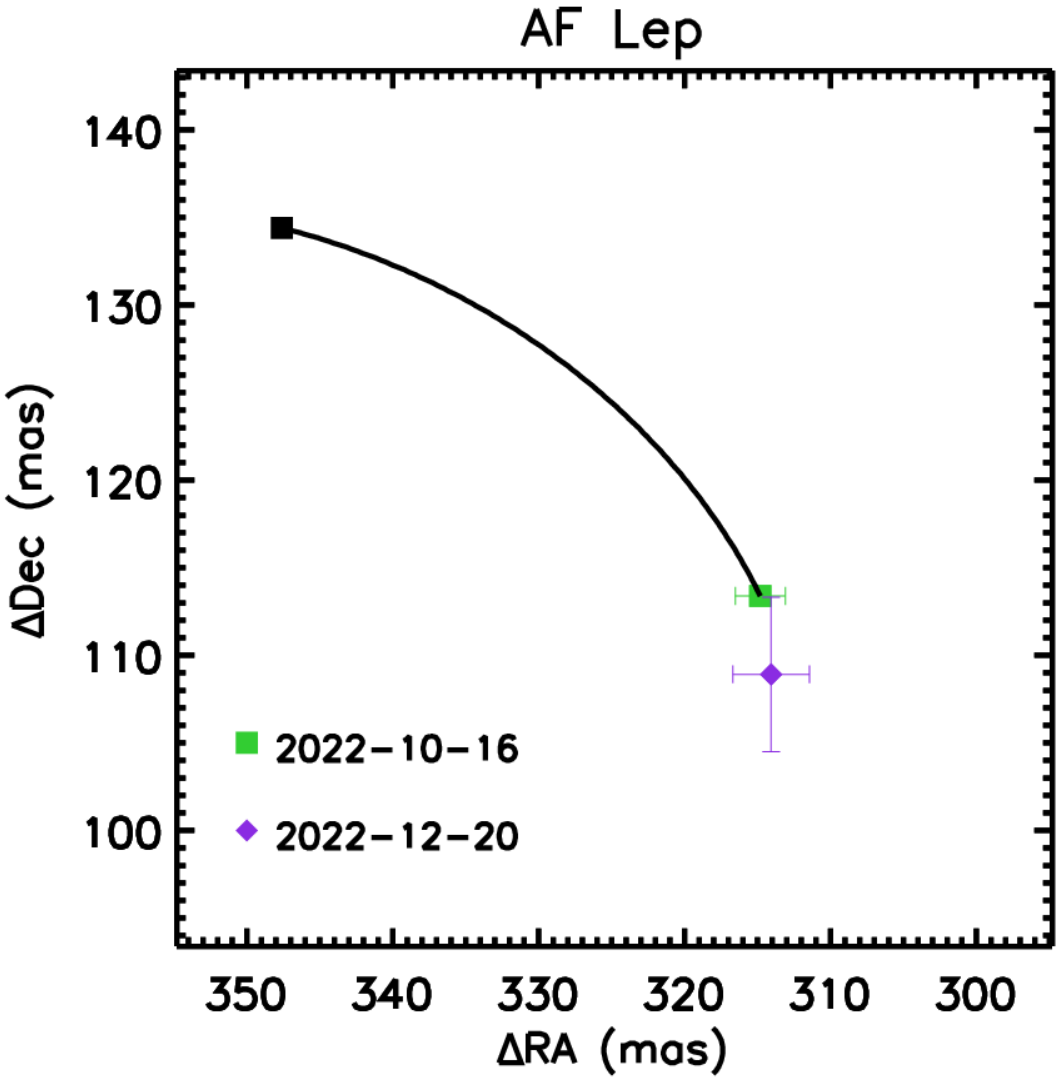}
\caption{Relative astrometry plot for the candidate companion of AF\,Lep. The green square represents the relative position of the companion at the first observing epoch while the violet diamond represents the relative position of the companion at the second epoch. The solid black line represents the expected motion relative to the host star of a background object while the black square is the expected position for a background object at the epoch of the second observation.}
\label{f:astroplot}
\end{figure}

\begin{figure}
\centering
\includegraphics[width=0.9\columnwidth]{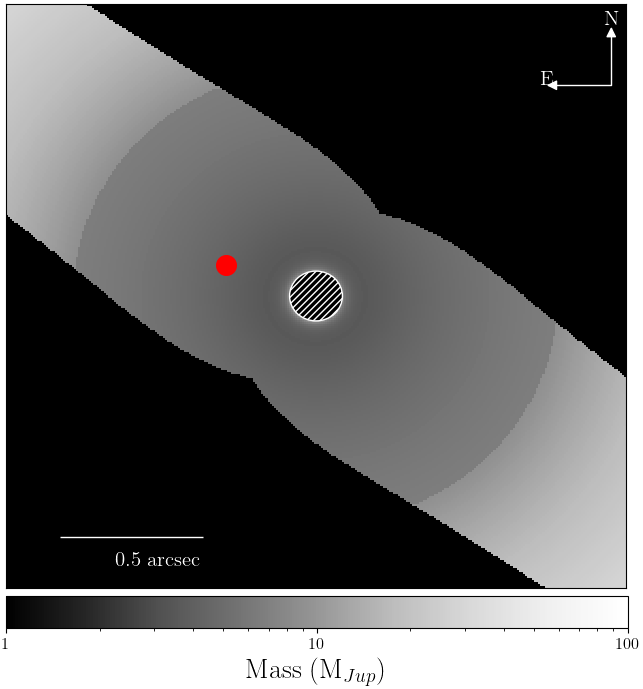}
\caption{2D maps, obtained with \textsc{FORECAST} showing in grey the sky area compatible with the PMa reported in Table~\ref{t:target}. The intensity of the grey areas changes according to the dynamical mass (in \MJup) responsible for the PMa at a given distance with the scale given by the lower color
bar. The position of AF\,Lep\,b for the first SPHERE epoch is shown as a red circle. 
 To avoid overlapping between the two positions we did not plot the position of the companion in the second epoch.
}
\label{f:probmap}
\end{figure}


\begin{table}
  \caption{IFS and IRDIS photometry of AF\,Lep\,b from both observing epochs.}\label{t:photo}
\centering
\begin{tabular}{ccccc}
\hline\hline
Sp. band &  Date & $\Delta$mag & Abs. mag. & Mag error \\
\hline
Y   & 2022-10-16 & --  & --  &  -- \\
J   & 2022-10-16 & 14.38  & 17.50   & 0.77 \\
H   & 2022-10-16 & 13.88  & 16.82   & 0.61 \\
K1  & 2022-10-16 & 11.56  & 14.34   & 0.26 \\
K2  & 2022-10-16 & 11.88  & 14.66   & 0.55 \\

Y  & 2022-12-20 & 14.09  & 17.21  & 0.61 \\
J  & 2022-12-20 & 13.95  & 17.07  & 0.33  \\
H  & 2022-12-20 & 13.52  & 16.46  & 0.36 \\
K1 & 2022-12-20 & 11.70  & 14.48  & 0.07 \\
K2 & 2022-12-20 & 11.79  & 14.57  & 0.07 \\
\hline\hline
\end{tabular}
\end{table}

\begin{table}
  \caption{Mass estimates for AF\,Lep\,b obtained from the photometry listed
    in Table~\ref{t:astro} and adopting the AMES-COND atmospheric models}
  \label{t:mass}
\centering
\begin{tabular}{cccc}
\hline\hline
 Sp. band &  Date & Mass (\MJup) & Err. Mass (\MJup) \\
\hline
Y  & 2022-10-16 &  --  & --  \\
J  & 2022-10-16 &  2.19  & 0.14  \\
H  & 2022-10-16 &  2.37  & 0.19  \\
K1 & 2022-10-16 &  4.14  & 0.34  \\
K2 & 2022-10-16 &  5.68  & 0.40  \\
Y  & 2022-12-20 &  2.76  & 0.20  \\
J  & 2022-12-20 &  2.40  & 0.19  \\
H  & 2022-12-20 &  2.72  & 0.19  \\
K1 & 2022-12-20 &  4.02  & 0.28  \\
K2 & 2022-12-20 &  5.80  & 0.41  \\
\hline
\end{tabular}
\end{table}

\begin{figure}
\centering
\includegraphics[width=0.9\columnwidth]{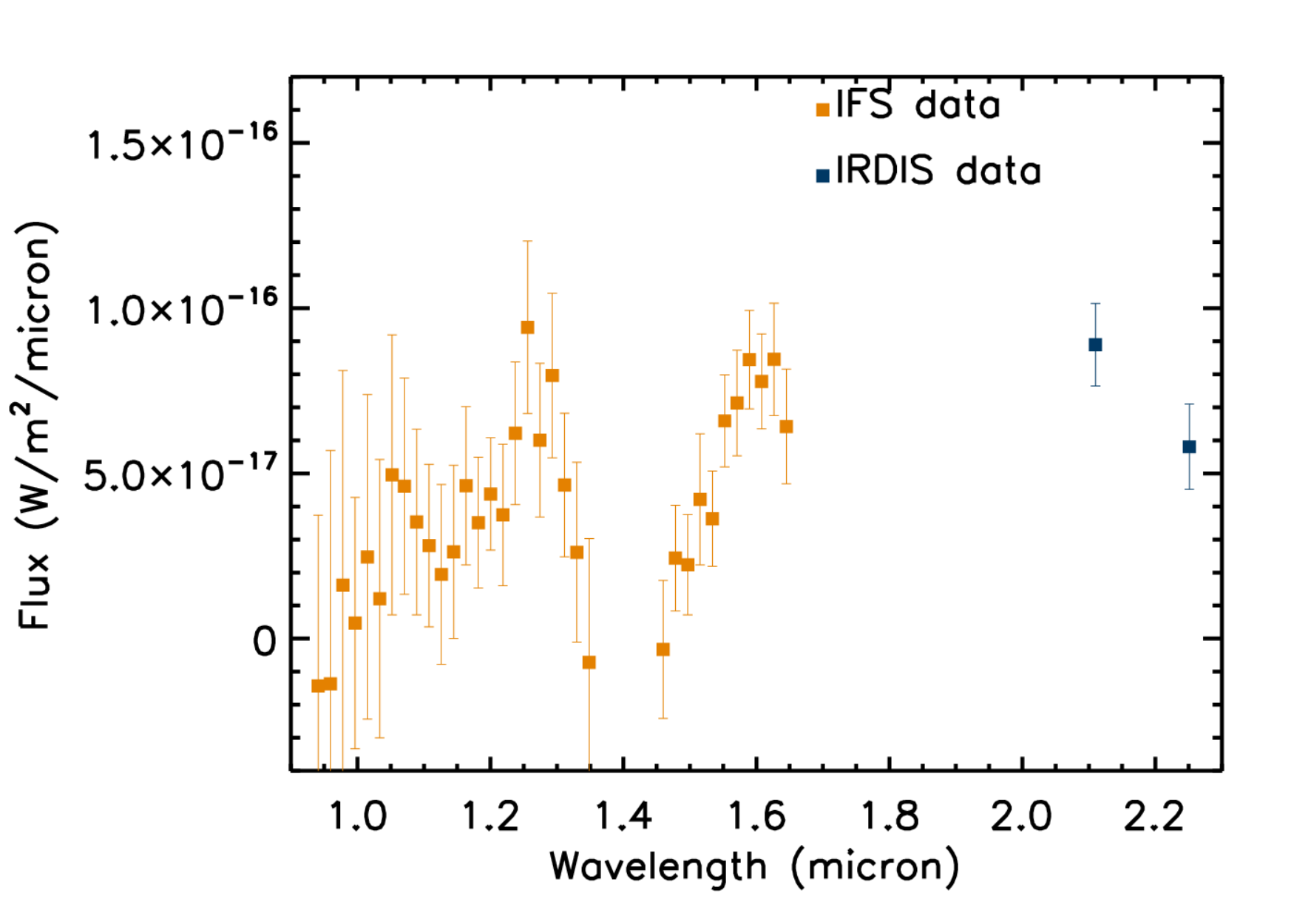}
\caption{Extracted spectrum for AF\,Lep\,b using both IFS and IRDIS data. Each data point is obtained by making a mean between the values
obtained in the two observing epochs. The orange squares are the IFS data points while the blue ones are the IRDIS data points.}
\label{f:spectrum}
\end{figure}

Figure~\ref{f:snrmap} shows the S/N maps obtained for IFS (left panel) and IRDIS (right panel) for each epoch. 
A point source is visible North-East of the star in both epochs, with a S/N of 8.6 and 9.1 for the IFS detection and 5.1 and 16.1 for IRDIS. \par
The astrometry of the companion was obtained using the negative planet method \citep[see 
e.g.,][]{2011A&A...528L..15B, 2014A&A...572A..85Z} and is reported in Table~\ref{t:astro} for both 
epochs. The astrometric calibration including the definition of the star center, the pixel scale, 
and angle of true north (TN) together with their respective errors are obtained following the strategy 
devised in \citet{2016SPIE.9908E..34M}\par 
Figure~\ref{f:astroplot} shows the relative astrometry of the companion candidate. 
The comparison between the expected position in the second epoch of a background object, represented by a black square, and the measured position of the candidate companion, represented by a violet diamond, clearly shows that the candidate is co-moving with AF\,Lep. 
Only the IFS data were used, due to their higher precision compared to the IRDIS values \citep{2014A&A...572A..85Z}, which does however lead to similar results. 
The large shift of the foreseen position for a background object between the two epochs is
mainly due to a favorable combination of the contributions from proper motion and parallactic motion. In this particular case the contribution of the latter is more important than that from proper motion.
\par
The IFS projected angular separation of AF\,Lep\,b corresponds to a projected physical separation of $9.0 \pm 0.1$~au in the first epoch and of $8.9 \pm 0.1$~au in the second epoch. This makes this planet one of the nearest to its host star among those discovered through direct imaging, with a semi-major axis very similar to the one of $\beta$\,Pic\,b \citep[see e.g.,][]{2021AJ....161..179B}. \par
The analysis of the measured PMa signal, combined with the information on the astrometry from the SPHERE data can be used to obtain further confirmation of the co-moving and planetary-mass nature of the detected companion. Using the \textsc{FORECAST} algorithm (Finely Optimised REtrieval of Companions of Accelerating STars\footnote{\url{https://maps.exoplanetsforecast.com/}}, see \citet{2022MNRAS.513.5588B} for a detailed description), we were able to identify the region where a companion compatible with the measured PMa should appear in the IFS FOV,  based on the PA of
the PMa vector that is reported in Table~\ref{s:target}.  
\textsc{FORECAST} evaluates the position angle of each pixel in the IFS image with respect to the center and then compares it with the measured value of the PA, identifying an optimal region on the image also taking into account the possible orbital motion of a companion at that position between the SPHERE observation and the Gaia observations. Finally, it also associates to each point of the resulting 2D map a value of the companion mass based on the PMa absolute value, calculated using the approach from \citet{2019A&A...623A..72K}. The resulting 2D map is shown in Figure~\ref{f:probmap}, with the position of the companion, perfectly within the optimal region, shown as a red circle, corresponding to a dynamical mass of $\sim$5.5~\MJup.  \par
The negative planet method was also used to extract the photometry of the companion from 
each wavelength channel of both IFS and IRDIS. In the first epoch, the IFS channels in Y spectral
band are largely dominated by the noise and it was not possible to extract any reliable results. 
Anyway it was possible to obtain reliable photometric results in J, H, K1 and K2 spectral bands in
the first epoch and for all the spectral bands during of the second epoch.
These results are listed in Table~\ref{t:photo}. The extracted photometry was then used to obtain a 
first estimate of the companion's mass, using the AMES-COND atmospheric models 
\citep{2003IAUS..211..325A}. The resulting values are listed in Table~\ref{t:mass} and range between 
$\sim$2 and $\sim$5.5~\MJup, placing AF\,Lep\,b among the least massive direct imaging planets 
detected so far. Using the same models we can also estimate the radius of the 
companion obtaining values ranging between 1.30 and 1.35~\RJup. \par
Using the same procedure used for the photometry of the companion we were also able to extract a low-resolution spectrum combining IFS (orange squares) and IRDIS (blue squares) data, making a weighted mean over the two epochs. The resulting spectrum is shown in Figure~\ref{f:spectrum}. 
The results from the two epochs are in good agreement for the H and the K
spectral bands while the agreement is less good for the Y and J spectral bands. This is due to the fact that the Y band data for the first epoch is completely dominated by noise while this was not true in the second epoch. 
The difference is less important in the J band, however the peak clearly visible in both epochs has an intensity in the second epoch that is just the 65\% in flux with respect to that in the first epoch. 
For this reason, the result in these spectral bands should be taken with some care. Finally, the spectral region around 1.4~\mic, affected by the presence of water telluric absorption, was also excluded from the extracted spectrum. \par





\section{Discussion}
\label{s:discussion}
In Section~\ref{s:discussion} we have presented the detection of a low mass candidate companion in both our SPHERE observations of AF\,Lep, retrieved with a S/N above 8-9 in the IFS data and a S/N of just 5 in the first epoch and of 16.1 in the second one in the IRDIS data. 
Although the conditions were not optimal in either epochs, the characteristics of the candidate, combined with the additional information provided by the measured Hipparcos-Gaia eDR3 PMa are such that we are confident that the detection is genuine. 

Indeed, the separation and the position angle for AF\,Lep\,b are coherent in all the observations 
so there are no doubts that we are observing a real object and not a badly subtracted speckle. The
differences in the position of the companion between the two epochs could be regarded as orbital motion 
even if it is difficult to draw a definitive conclusion on this because of the small time
span between the two observations. The comparison between the astrometric position of the
companion in the two epochs is a clear confirmation that it is actually gravitationally bound to the 
star and not a background object. In this case, moreover, we have available also the astrometric measures and the PMa signal obtained by \citet{2022A&A...657A...7K} to further confirm that it 
is actually a bound companion.
It is important to note that the position of the detected candidate
is coherent with the expected position for the object causing the
PMa signal within the uncertainties related to the time elapsed
between the Gaia and SPHERE epochs. 
Furthermore, the photometric mass obtained for our candidate is coherent with the
dynamical mass expected at that separation to justify the PMa
signal. Further follow-up observations will be important to further precise its orbital and physical characteristics. However, the
convergence of different detection methods towards results that
are in good agreement each other allows us to conclude that the
detected companion is actually a bound object. \par
There are more reasons to exclude that the the candidate
companion is a background object. First of all, the candidate
companion is very red and it has a spectrum compatible with
being a small mass physical companion to AF\,Lep. Secondly,
given the large proper motion of the star (about 50 mas/a), if the
candidate companion were a background object it would have
been about 0.9\as South-East of the star at the epoch of the first
observation of AF\,Lep reported by \citet{2016A&A...594A..63G} in
February 2006. At this separation the limiting contrast of their
observation is 13 mag (in the CH4 band) and the candidate companion should 
have been detected, while it was not. In addition,
there are 760 background stars in Gaia DR3 with a contrast of
< 15 mag in the G-band within 10 arcmin from AF\,Lep. This
contrast is comparable to the observed contrast of the candidate
companion, that should likely be an M-star if a background object. The surface density of background stars is then $6.72\times10^{-4}$
stars/$arcsec^2$. The probability of finding one of them at 0.335\as
or less from AF\,Lep is then $2.3\times10^{-4}$. From all these considerations we can then conclude that it is very unlikely that the candidate companion is a background object. \par
This target has been observed in several high-contrast imaging survey. We can consider then why it 
was never detected before. The detection of AF\,Lep\,b on our images is due to a
combination of the very good sensitivity of SPHERE at short
separations and of the epoch when this observation was done.
Indeed, given the position of AF\,Lep\,b and the direction of
the PMa vector ($PA=233.5^{\circ}$) and considering the fact that it is almost parallel to the companion's PA ($\sim$$70^{\circ}$) suggesting an high orbital inclination, the separation between the star and the planet was much smaller in the recent past. Given the
absolute value of the PMa and the mass ratio we determined
(q$\sim$0.005), we may roughly estimate that the speed of the planet
around the star is currently $\sim$53~mas/a. The planet should then
have been very close to the star (separation$<$0.1\as) at the epochs
of the most recent observations of AF\,Lep like ISPY \citep{2020A&A...635A.162L}, 
GPI \citep{2019AJ....158...13N}, LEECH \citep{2018AJ....156..286S} that were performed between 
2014 and 2016. Older observations like SPOTS \citep{2018A&A...619A..43A}, NICI 
\citep{2013ApJ...777..160B}, the International Deep planet survey \citep{2016A&A...594A..63G} and
\citet{2007A&A...472..321K} were probably not deep enough to detect this faint object at the small 
separation it lies. 

In the following sections we discuss in detail the results obtained for AF\,Lep\,b, together with their implication on the nature of this object. 

\subsection{Spectral fit}
\label{s:spectrumfit}

\begin{figure}
\centering
\includegraphics[width=0.9\columnwidth]{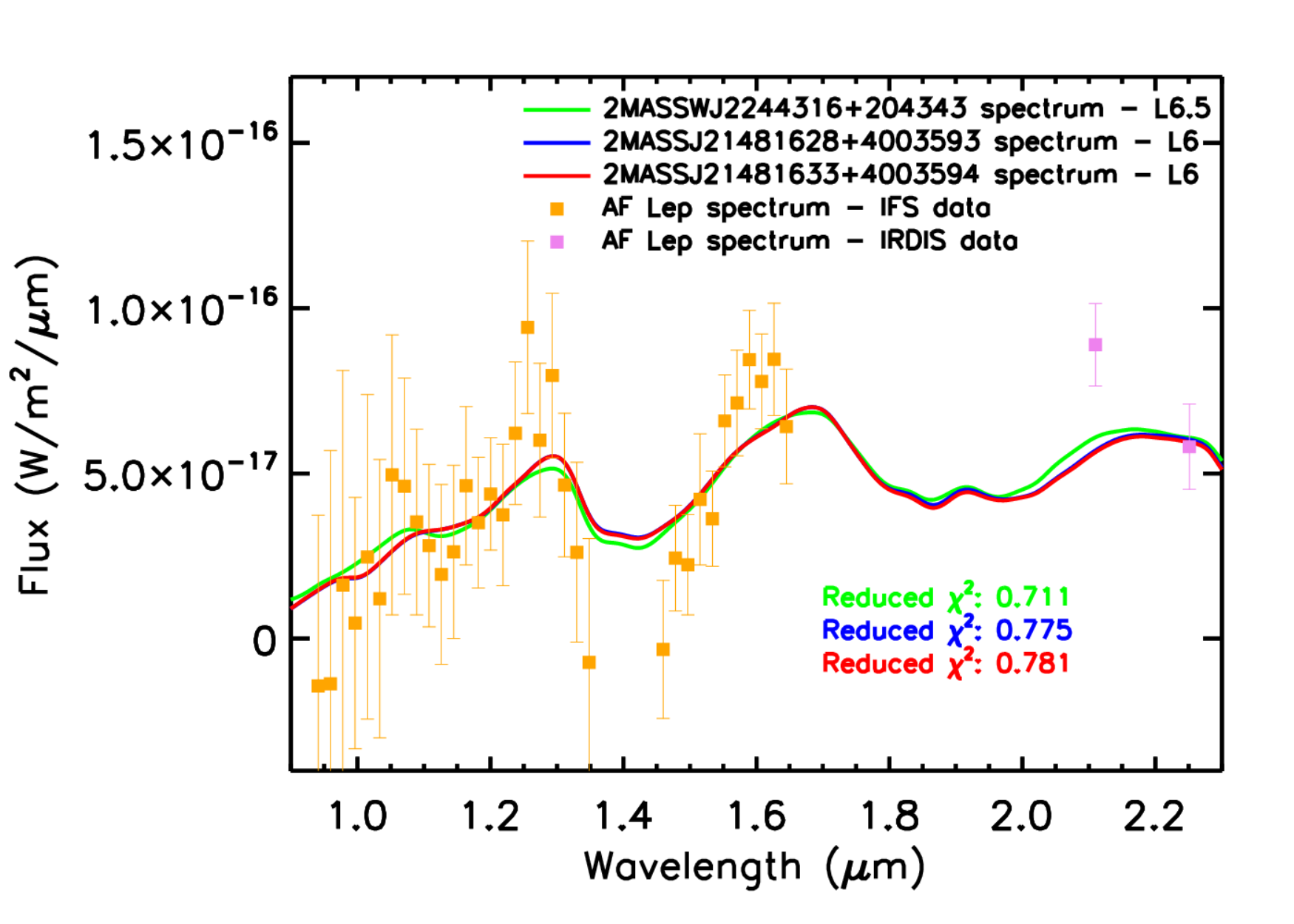}
\caption{Comparison of the extracted spectrum of AF\,Lep\,b with the three best-fit template spectra from the SpexPrism spectral library. The IFS data points are 
orange squares while the IRDIS data points are violet squares. The best three template spectra are represented by the green, blue and red solid lines. They are partially overlapped each other. The reduced $\chi^2$ for each template spectrum is reported with the corresponding color. } 
\label{f:templatefit}
\end{figure}

\begin{figure}
\centering
\includegraphics[width=0.9\columnwidth]{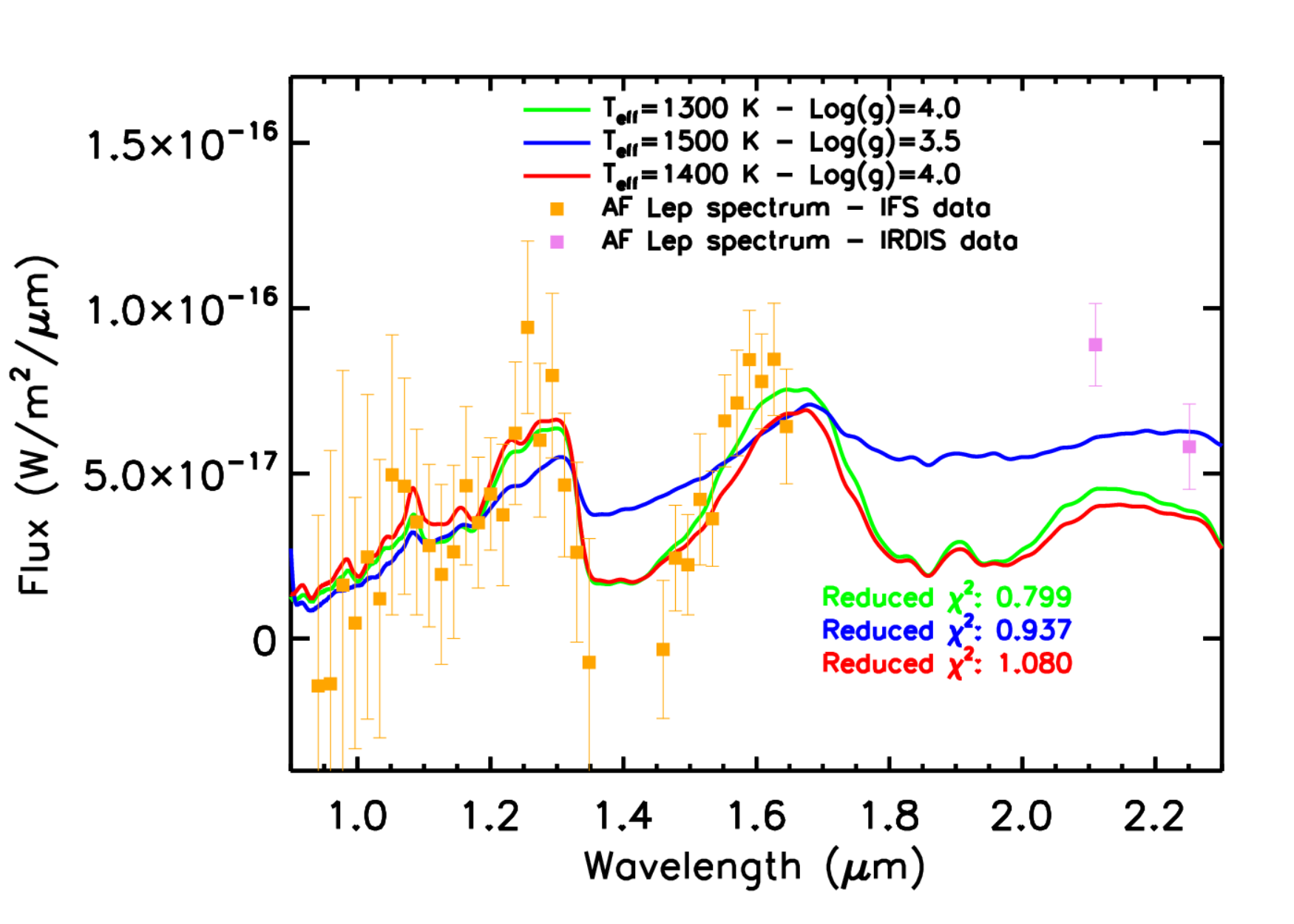}
\caption{Comparison of the extracted spectrum of AF\,Lep\,b with the three best fit BT-Settl atmospheric models. The IFS data points are 
orange squares while the IRDIS data points are violet squares. The best three atmospheric models are represented by the green, blue and red solid lines. 
The reduced $\chi^2$ for each atmospheric model is reported with the corresponding color.}
\label{f:modelfit}
\end{figure}

To obtain some information on the physical characteristics of AF\,Lep\,b we fitted its spectrum 
extracted as described in Section~\ref{s:res} both to template spectra and 
atmospheric models. 
We stress here that these results present a large uncertainty and they should be regarded with some care because of the large error bars for each spectral data point as can be seen in Figure~\ref{f:spectrum}. 
We obtained the template spectra from the SpexPrism spectral Library \citep{2014ASInC..11....7B} and the results of this fitting procedure can be seen in Figure~\ref{f:templatefit} where we compare the extracted spectrum of AF\,Lep\,b with that of the three best-fit template spectra. 

The best fit is obtained with $\sim$L6 spectral types. Anyhow, comparably good fits are obtained for other later L spectra types. We can then conservatively conclude that AF\,Lep\,b present a late-L spectral type. \par As atmospheric models we adopted the BT-Settl models \citep{2014IAUS..299..271A}. The results are shown in Figure~\ref{f:modelfit} where we compare the extracted spectrum of AF\,Lep\,b with those of the three best-fit atmospheric models. In this case, the best fits are obtained for models with $T_\mathrm{eff}$ between 1300 and 1500~K and surface gravity of $\logg\sim$3.5-4.0~dex. Also in this case however a larger number of models at different $T_\mathrm{eff}$ give a comparably good fit. We can then assume a quite large $T_\mathrm{eff}$ range between 1000 and 1700~K. 

\subsection{PMa analysis}
\label{s:pmaana}

\begin{figure}
\centering
\includegraphics[width=0.9\columnwidth]{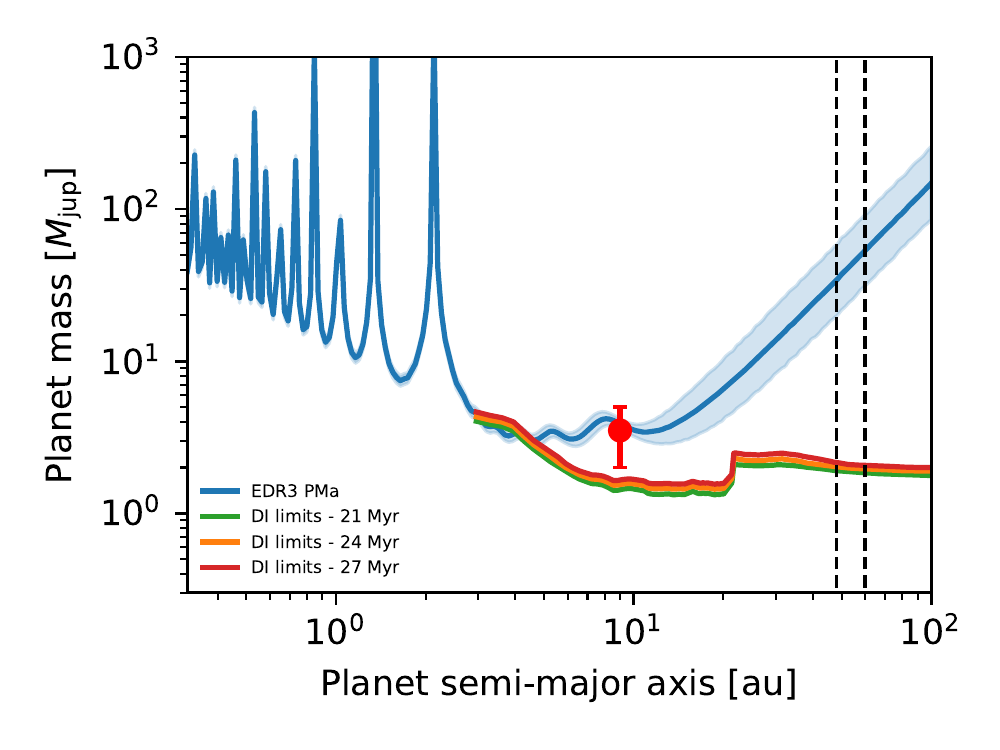}
\caption{Comparison of the direct imaging results obtained from the SPHERE observations with the PMa results. The solid blue line represents the mass of the object causing the
PMa signal as a function of the separation from the host star. The shaded blue area represents the $1\sigma$ confidence interval for this measure. The mass limits obtained from the SPHERE 
observations for the minimum, expected and maximum ages are represented by the green, orange, and red solid lines, respectively. The red circle represents the position on this diagram
of AF\,Lep\,b. The error bar in separation is too small to be shown in this Figure. The two vertical black dashed lines represent the position of the external belt of the debris
disk hosted by AF\,Lep.}
\label{f:pmacomp}
\end{figure}

An alternative way to visualise the results obtained until now this is provided in 
Figure~\ref{f:pmacomp}, which shows the estimated mass of the object causing the PMa signal as a 
function of the separation from the host star (solid blue line), derived once again following the 
approach by \citet{2019A&A...623A..72K}. 
The mass limits obtained from the SPHERE direct imaging data are also shown for different ages of the 
system, highlighting that the SPHERE images would have allowed for the detection of any companion 
compatible with PMa at separations lower than 4~au. 
The position of AF\,Lep\,b is marked by a red circle, showing a very good agreement between the mass 
obtained from the photometry and the PMa dynamical mass prediction at the companion's separation. 


\subsection{Orbital fit and dynamical mass estimate}
 \label{s:disc_astro}
\begin{figure*}
     \centering
     \includegraphics[width=\textwidth]{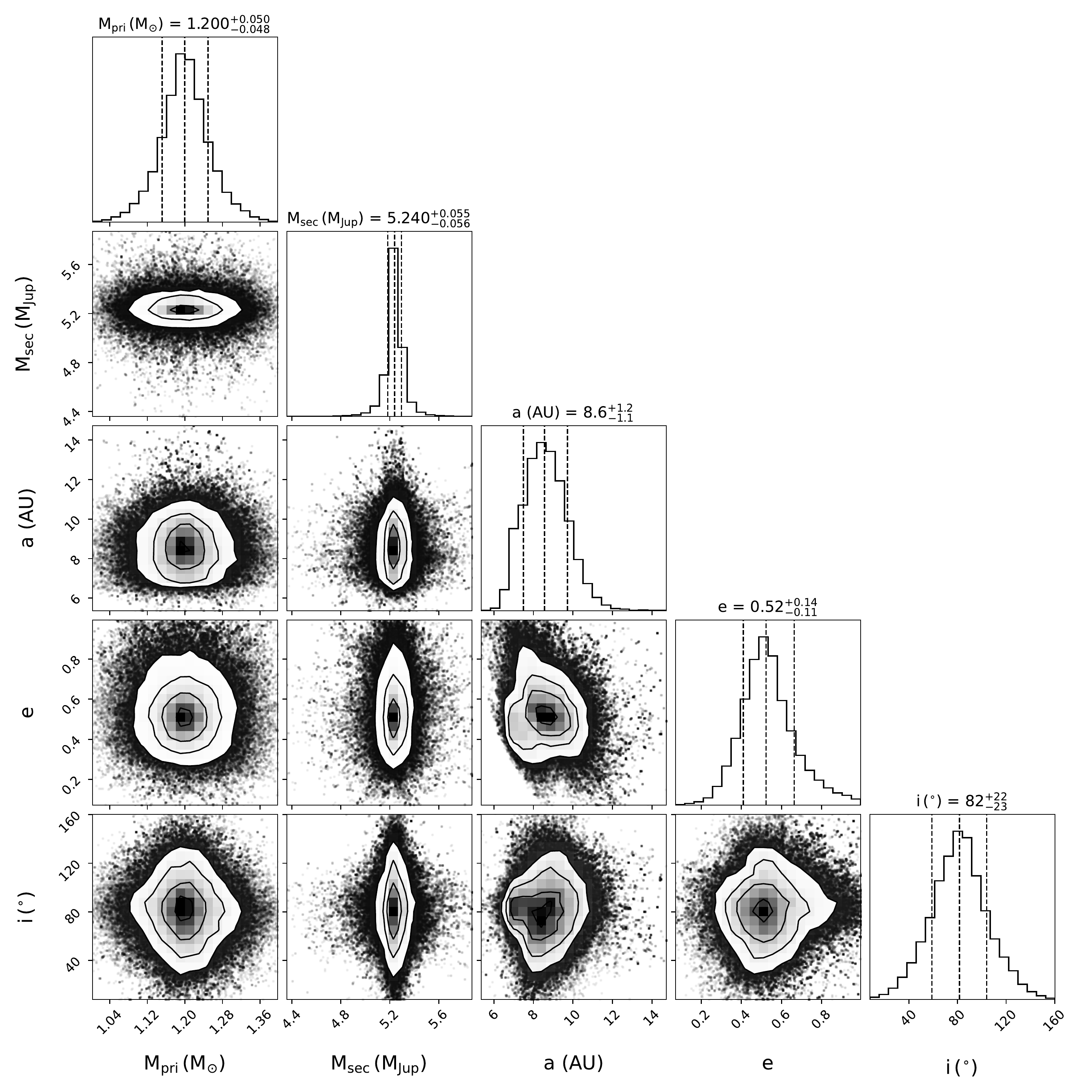}
     \caption{Corner plot showing the results of the orbital fit performed with the orvara tool. The best-fit values are reported in Table~\ref{t:companion}. }
     \label{f:orvara_fit}
 \end{figure*}

We used the orvara\footnote{\url{https://github.com/t-brandt/orvara}}
\citep[Orbits from Radial Velocity, Absolute, and/or Relative Astrometry;][]{Brandt:2021AJ} tool to combine the SPHERE astrometry for our two epochs and the PMa information and perform a first orbital fit for AF\,Lep\,b and obtain a more precise estimate of its dynamical mass. 
 The results of the fit are summarised by the corner plot shown in Figure~\ref{f:orvara_fit} and the best-fit values are reported in Table~\ref{t:companion}, together with the companion's physical parameters derived from the spectral fit in Section~\ref{s:spectrumfit}. \par
Overall the results of the fit agree with the estimates of the mass and orbital parameters discussed in the previous Section. In particular the inclination is indeed rather high, corroborating the discussion about the previous non-detections, and the mass agrees with both the value obtained from the photometry and the one resulting from the PMa-only analysis. This once again reinforce the argument towards the genuine co-moving nature of the companion. 

\renewcommand{\arraystretch}{1.5}
\begin{table}
\caption{Summary of the physical and orbital parameters of AF\,Lep\,b.}
\label{t:companion}
\centering
\begin{tabular}{ll}
\hline \hline
Parameter                       & Value         \\
\hline
Temperature (K)                 & 1000-1700  \\
Spectral Type                   & Late L ($>$L6) \\
\hline
Photometric Mass  (\MJup)               & 2-5.5\\ 
Dynamical Mass from PMa     (\MJup)     & $\sim$5.5\\
Dynamical Mass from orvara  (\MJup)     & ${5.237}_{-0.10}^{+0.085}$ \\ 
Primary Mass from orvara    (\MSun)    & ${1.201}_{-0.056}^{+0.058}$ \\
Mass ratio                      & ${0.00416}_{-0.00021}^{+0.00022}$\\
\hline
Projected separation (arcsec)   & 0.334$\pm$0.001\\
Position angle (deg)            & 70.65$\pm$0.26 \\
Inclination (deg)          & $82_{-23}^{+22}$  \\
Semi-major axis (au)            & ${7.99}_{-0.92}^{+0.85}$\\
Ascending node (deg)            & ${249}_{-12}^{+16}$\\
Eccentricity                    & ${0.47}_{-0.13}^{+0.17}$\\
Period (years)                    & ${20.6}_{-3.5}^{+3.4}$\\
Argument of periastron (deg)    & ${46.4}_{-8.9}^{+9.3}$\\
T$_{0}$ (JD)                         & ${2456328}_{-503}^{+1386}$\\
\hline \hline 
\end{tabular}
\end{table}

\section{Conclusion}
\label{s:conclusion}

We presented in this paper the near-infrared VLT/SPHERE observations of the nearby, young star AF\,Lep.
The main result of this work is the discovery of a planetary-mass companion at low separation from the star. The measured separation of the companion from the star is of $\sim$0.335\as corresponding to $\sim$9~au. We confirmed that it is gravitationally bound to its host star by comparing its position relative to the host star in two different epochs.
Moreover, further confirmation was possible thanks to astrometric considerations based on the PMa measured for the host star. 
 Using the orvara tool, we derived the orbital parameters compatible with the SPHERE astrometry at the two epochs, as well as with the direction and amplitude of the PMa signature. We were also able to derive a first estimate of the companion's dynamical mass, $\sim$5.2~\MJup, which is in good agreement with the value obtained from the analysis of the PMa alone (5.5~\MJup) as well as with the 2-5~\MJup value obtained from the SPHERE photometry, using atmospheric models.\par 
The S/N of the detection is generally quite low ranging between $\sim$5 and $\sim$16 according to the observing epoch and to the instrument used. This is especially true in the Y and J 
spectral bands for which we obtained large error bars. This partially hampers our capability to extract information from the low-resolution spectrum obtained from SPHERE.
In any case, we can conservatively conclude that AF\,Lep\,b is a late-L spectral type object with a $T_\mathrm{eff}$ ranging between 1000 and 1700~K. Observing this target in better
weather conditions could allow obtaining a detection with a higher S/N. As a consequence, this will allow to extract of a less noisy spectrum with the possibility to obtain more
precise physical characteristics. \par

Finally, it should be noted that AF\,Lep\,b is the first companion with a mass well below the deuterium-burning limit detected around a star showing an astrometric signature and among the smallest directly imaged companions overall. 
Most of the other directly imaged companions to accelerating stars in fact have masses predominantly in the brown dwarf regime \citep[see e.g. HIP\,21152\,B, HIP\,29724\,B, HIP\,60584\,B and HIP\,63734\,B][]{2022MNRAS.513.5588B}. The recently discovered planetary-mass companions to HIP\,99770 \citep{2022arXiv221200034C} and HD\,206893 \citep[HD\,206893\,c;][]{2022arXiv220804867H} are still more massive than AF\,Lep\,b ($\sim$16~\MJup and $\sim$12.7~\MJup, respectively). \par
Given its mass and separation, and the host star spectral type, AF\,Lep\,b can be considered the first directly detected Jupiter analogue orbiting an accelerating star.  
Its resemblance to the Solar System, including the presence of a debris belt at a separation up to $\sim$60~au resembling the Kuiper belt, makes the AF\,Lep system an ideal target for future in-depth characterisation, both with GRAVITY or JWST to further refine physical and orbital characteristics of AF\,Lep\,b, and with future more precise instruments to search for additional companions.

\begin{acknowledgements}
This work has made use of the SPHERE Data Center, jointly operated by
OSUG/IPAG (Grenoble), PYTHEAS/LAM/CeSAM (Marseille), OCA/Lagrange (Nice) and
Observatoire de Paris/LESIA (Paris). \par
This work has made use of data from the European Space Agency (ESA) mission
{\it Gaia} (\url{https://www.cosmos.esa.int/gaia}), processed by
the {\it Gaia} Data Processing and Analysis Consortium (DPAC,
\url{https://www.cosmos.esa.int/web/gaia/dpac/consortium}). Funding for
the DPAC has been provided by national institutions, in particular, the
institutions participating in the {\it Gaia} Multilateral Agreement. \par
This research has made use of the SIMBAD database, operated at CDS,
Strasbourg, France. \par
D.M., R.G., and S.D. acknowledge the PRIN-INAF 2019 "Planetary systems at young ages (PLATEA)" and 
ASI-INAF agreement n.2018-16-HH.0. A.Z. acknowledges support from the FONDECYT Iniciaci\'on en investigaci\'on project number 11190837 and ANID -- Millennium Science Initiative Program -- Center Code NCN2021\_080. S.M.\ is supported by the Royal Society as a Royal Society University Research Fellow. \par
SPHERE is an instrument designed and built by a consortium consisting
of IPAG (Grenoble, France), MPIA (Heidelberg, Germany), LAM (Marseille,
France), LESIA (Paris, France), Laboratoire Lagrange (Nice, France),
INAF-Osservatorio di Padova (Italy), Observatoire de Gen\`eve (Switzerland),
ETH Zurich (Switzerland), NOVA (Netherlands), ONERA (France) and ASTRON
(Netherlands), in collaboration with ESO. SPHERE was funded by ESO, with
additional contributions from CNRS (France), MPIA (Germany), INAF (Italy),
FINES (Switzerland) and NOVA (Netherlands). SPHERE also received funding
from the European Commission Sixth and Seventh Framework Programmes as
part of the Optical Infrared Coordination Network for Astronomy (OPTICON)
under grant number RII3-Ct-2004-001566 for FP6 (2004-2008), grant number
226604 for FP7 (2009-2012) and grant number 312430 for FP7 (2013-2016).
For the purpose of open access, the authors have applied a Creative Commons Attribution (CC BY) licence to any Author Accepted Manuscript version arising from this submission. 

\end{acknowledgements}

\bibliographystyle{aa}
\bibliography{HIP25486}

\appendix

\end{document}